\documentclass[11pt,twoside]{article}
\usepackage{asp2010}

\resetcounters

\begin{document}

\title{Hall drift in the crust of neutron stars - necessary for radio pulsar activity?}
\author{Ulrich~Geppert,$^{1,2}$ Janusz~Gil,$^1$ Giorgi~Melikidze,$^1$ J.A.~Pons,$^3$ and D.~Vigan\`{o}$^3$
\affil{$^1$Kepler Institute of Astronomy, University of Zielona G\'{o}ra, Lubuska
    2, 65-265, Zielona G\'{o}ra, Poland}
\affil{$^2$ DLR-Institute of Space Systems,Bremen, Germany}
\affil{$^3$ Department of Applied Physics, University of Alicante, Alicante, Spain}
}

\begin{abstract}
    The radio pulsar models based on the existence of an inner accelerating gap
    located above the polar cap rely on the existence of a small scale, strong
    surface magnetic field $B_s$. This field exceeds the dipolar field $B_d$,
    responsible for the braking of the pulsar rotation, by at least one order
    of magnitude. Neither magnetospheric currents nor small scale field
    components generated during neutron star's birth can provide such field
    structures in old pulsars. While the former are too weak to create
    $B_s \gtrsim 5\times 10^{13}$G$\;\gg B_d$, the ohmic decay time of the
    latter is much shorter than $10^6$ years.\\
    We suggest that a large amount of magnetic energy is stored in a toroidal
    field component that is confined in deeper layers of the crust, where the
    ohmic decay time exceeds $10^7$ years. This toroidal field may be created
    by various processes acting early in a neutron star's life. The Hall drift
    is a non-linear mechanism that, due to the coupling between different
    components and scales, may be able to create the demanded strong, small
    scale, magnetic spots.\\
    Taking into account both realistic crustal microphysics and a minimal
    cooling scenario, we show that, in axial symmetry, these field structures
    are created on a Hall time scale of $10^3$-$10^4$ years. These magnetic
    spots can be long-lived, thereby fulfilling the pre-conditions for the
    appearance of the radio pulsar activity. Such magnetic structures created
    by the Hall drift are not static, and dynamical variations on the Hall time
    scale are expected in the polar cap region.
\end{abstract}

\section{The basic idea.}
        
The Partially Screened Gap Model relies on an intimate interplay of the
cohesive energy in the polar cap surface layer and the corresponding surface
temperature $T_s$ as well as on the partial screening by the thermal outflow
of iron ions  (\cite{2003_Gil}). Both quantities depend on the local surface field
strength $B_s$. The condition for the existence of an accelerating gap has
been calculated by  \cite{2007_Medin}. The balance of heating by the
bombardment with ultrarelativistic particles and cooling by radiation
returns for typical radio pulsar parameter $T_s \gtrsim 10^6$ K
\cite{2003_Gil}, a significantly higher value than the cooling age predicts.
In order to enable the creation of a gap for such high  $T_s$, $B_s$ has to
be larger than $5\times 10^{13}$ G, perhaps even larger than $10^{14}$ G.
Simultaneous X-ray and radio observations with X-ray spectra that can be fitted
by blackbody radiation (\cite{2006_Kargaltsev,2005_Zhang}) support these
estimates. Though these fits have to be considered with caution (see ERPM
talk of W. Hermsen) they may be indicating that the base of the open field
lines on the stellar surface (heated to temperatures above $10^6$ K) is much
smaller than the conventional polar cap (\cite{1975_Ruderman}). Flux conservation
arguments lead to
$B_s \gtrsim 5\times 10^{13}$ G $\gg B_d \sim 5\times 10^{12}$ G
(for a typical radio pulsar).

In order to allow an efficient electron-positron pair creation rate within the
accelerating gap, the curvature radius of the magnetic field lines must be
$R_{B_s} \ll R_{B_d} \sim 100$ km (\cite{1975_Ruderman}). This is valid when
either curvature radiation or inverse Compton scattering are the dominating
processes (\cite{2000_Melikidze},\cite{2011_Szary}).
Polar cap surface fields of the required strength and curvature cannot be
present since the birth of the neutron star, because the electric conductivity
during the first $\sim 10^4$ yr is relatively low and, for small-scale
structures is $\lesssim 1$ km, the ohmic decay time in the subsurface crustal
layers is typically only a few $10^2$ -  a few $10^3$ yrs. Therefore, the
demanded $B_s$ has to be (re-)created and maintained over the lifetime of
radio pulsars, i.e. $\sim 10^6-10^7$ yr.
Therefore, there must be a large reservoir of magnetic energy, stored in
regions where it can survive for $\gtrsim 10^6$ yr, which, at some point over
the lifetime of radio pulsars, can be tapped for forming this $B_s$.

Since magnetospheric currents are not a plausible mechanism to create the
demanded $B_s$ - structures (\cite{2001_Hibschman}), the Hall drift of the
crustal magnetic field turns out to be a possible alternative to
explain the existence of the of small scale, strong surface fields.  We
propose that the energy reservoir is a large scale crustal toroidal field
whose maintaining currents circulate in deeper layers, where the high electric
conductivity ensures a sufficiently long lifetime. Due to the non-linear
interaction of the crustal and/or core based poloidal field $\sim B_d$ with
the toroidal crustal field, a magnetic spot in the vicinity of the polar cap
can be created.

\section{Results from simulations.}

\begin{figure}[!ht]
    \centering
    \includegraphics[width=4.cm]{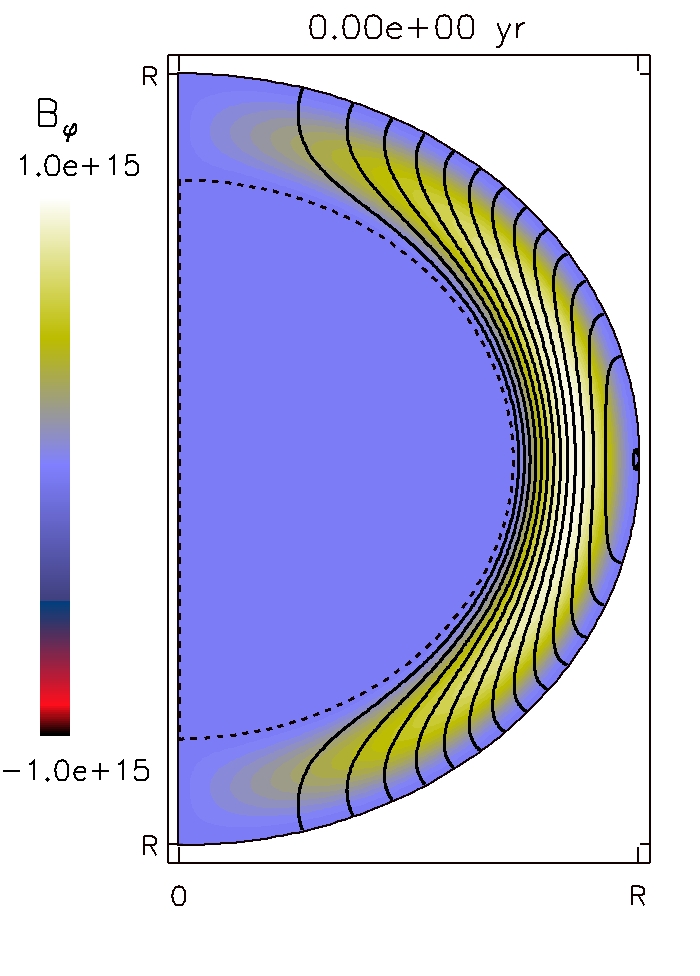}
    \includegraphics[width=4.cm]{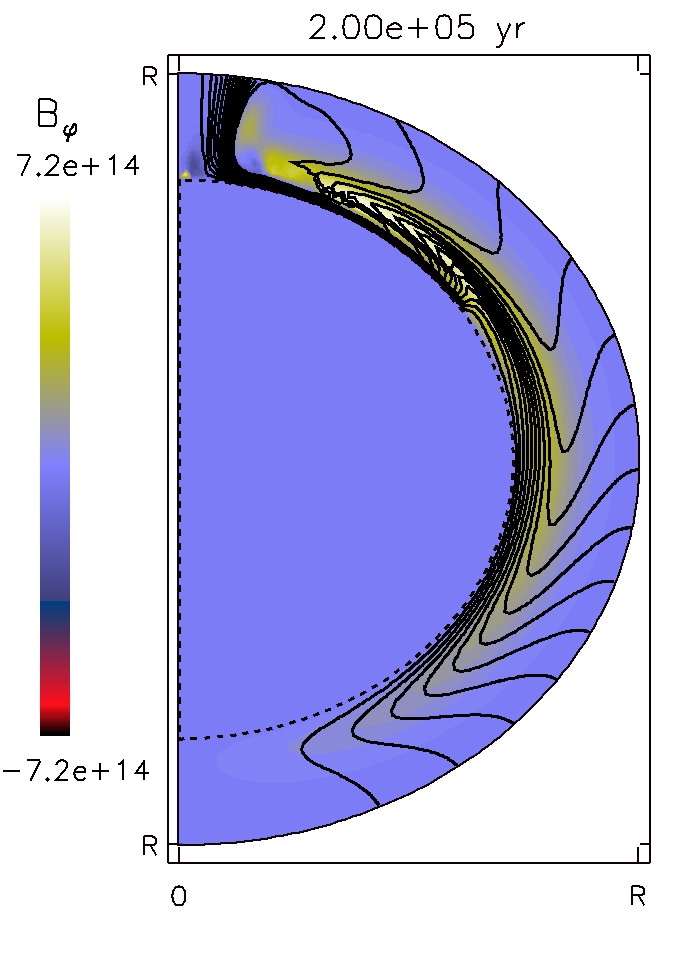}
    \includegraphics[width=4.cm]{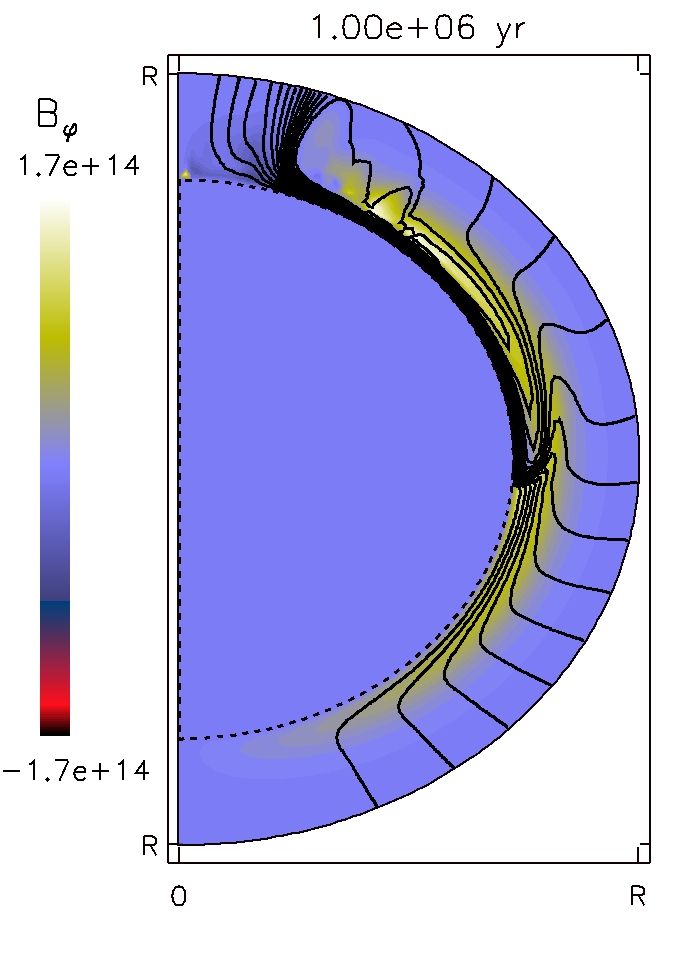}
    \caption{Structure of the crustal magnetic field at $t=0$ (left), after
        $2\times 10^5$ (middle) and after $10^6$ yrs (right). The poloidal
        field is shown by solid lines, the isolines of the toroidal field are
        color coded. The crustal region has been stretched a by factor of 4
        for visualization purposes. The complete movie showing the field
        dynamics is available at {\it http://personal.ua.es/en/daniele-vigano/hall-pulsar.html}}
\label{Bcrust_t}
\end{figure}

The evolution of the magnetic field in the crust, where electrons are the only
carriers of the field generating electric currents, is described by the Hall
induction equation. For details see \cite{2007_Pons}.
The Hall drift can generate very small scale structures, such as current
sheets and shock-like patterns, out of a large scale field. A numerical code
based on a finite difference scheme and non-local boundary conditions
(\cite{2012_Vigano}) has been used to follow the evolution of the magnetic field
under typical conditions, with realistic microphysics (\cite{2008_Aguilera}) and
for the minimal cooling scenario (\cite{2004_Page}).


\begin{figure}[!ht]
    \centering
    \includegraphics[width=9cm]{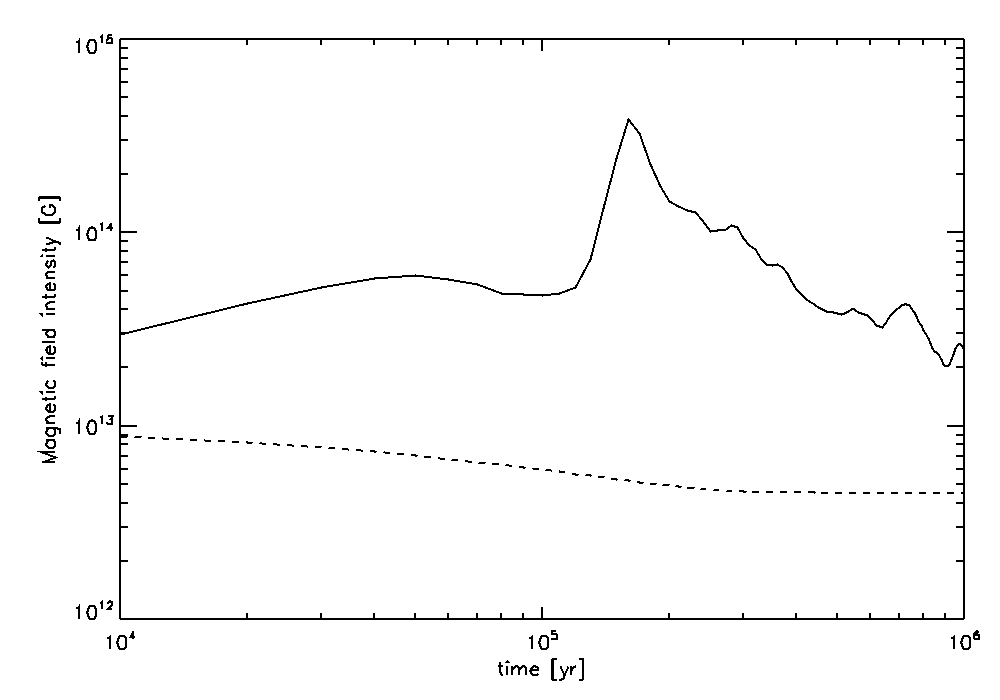}\\
    \includegraphics[width=9cm]{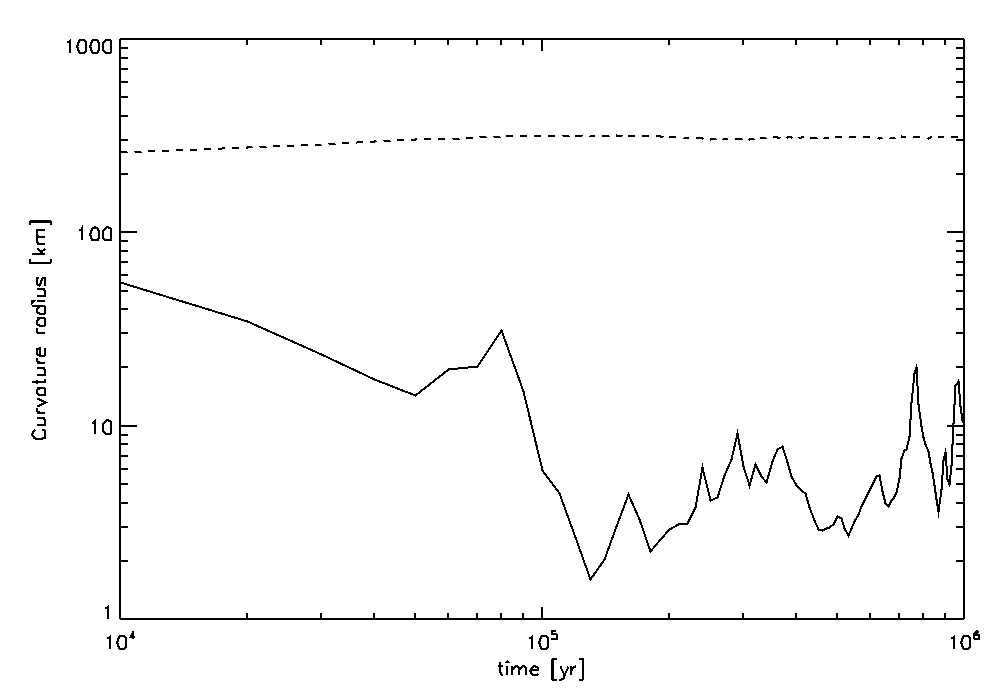}
    \caption{Temporal evolution of $B_s$ (left) and $R_{B_s}$ (right) near the
    north pole, considering only Ohmic dissipation (dashed) or including the
    Hall term (solid). We show averages of the numerical values of $B_s$ and
    the minimum of $R_{B_s}$ in the region $1^{\circ}-5^{\circ}$ from the
    north pole.}
    \label{B_RC_t}
\end{figure}

We assume an initial field configuration as depicted in the left panel of
Fig.~\ref{Bcrust_t}.
MHD equilibria suggest that the toroidal field is concentrated in an equatorial
belt, filling only a small part of the crust volume. However, there are
mechanisms conceivable (e.g. small scale dynamos, magneto-rotational and
thermoelectric instabilities) that can create strong internal fields soon
after a neutron star's birth. In some cases, the crustal field can have a
large toroidal component that fills most of the crust volume, which results
in a strongly peaked temperature distribution, consistent with the observed
large pulsed fraction of Kes 79 (\cite{2012_Shabaltas}). 

Fig.~\ref{Bcrust_t} shows the structure of the crustal field after
$2\times 10^5$ and $10^6$ yrs. The strong, small scale field structures are
formed within a few $10^4$ yr and can survive over a million years because of
the relatively low temperature (high conductivity), in contrast with the fast
dissipation of initial structures in very young, hot neutron stars. The
temporal evolution of $B_s$ and $R_{B_s}$ near the north pole is shown in
Fig.~\ref{B_RC_t}. After $\sim 10^5$ yrs $B_s \gtrsim 10\times B_d$ there;
the radius of field line curvature is about two orders of magnitude smaller
than in case of a purely resistive field evolution. Obviously, the Hall drift
introduces another time scale into the polar cap dynamics that is neither
determined by the pulsar rotation nor by the $\vec{E} \times \vec{B}$-drift
but solely by the non-linear evolution of the crustal magnetic field. We must
note that the large magnetic Reynolds numbers close to the polar surface
results in some numerical noise that makes us consider the absolute values
of $B_s$ and $R_{B_s}$ with caution.

\section{Conclusion}

Our main conclusion is, therefore, that the Hall drift is a viable process,
that might create both on a correct time scale and on proper scale lengths
the surface magnetic field configurations that enable a neutron star to appear
as radio pulsar.
Although the model is limited to the 2D, axially symmetric case, so that no
"real" spots, 
limited both in meridional and azimuthal direction, can arise, the results
are promising and should motivate further investigations in this field. 

\bibliography{bibliography}
    

\end{document}